\documentclass[12pt,a4,paper]{article}
\usepackage{graphicx}
\title{Several results from numerical investigation of nonlinear waves connected to blood flow in an elastic tube of variable radius}
\author{Zlatinka I. Dimitrova}
\date{"G. Nadjakov" Insitute of Solid State Physics, 72 Tzarigradsko Chaussee 72 Blvd., 1784 Sofia, Bulgaria\\
e-mail: zdim@issp.bas.bg}

\begin{document}
\maketitle
\begin{abstract}
We investigate flow of incompressible fluid in a cylindrical tube with elastic 
walls. The radius of the tube may change along its length. The discussed problem
is connected to the  blood flow in large human arteries and especially to 
nonlinear wave propagation due to the pulsations of the heart. The long-wave
approximation for modeling of waves in blood is applied. The obtained model
Korteweg-deVries equation possessing a variable coefficient is reduced to a nonlinear dynamical system of 3 first order differential equations. The
low probability of arising of a solitary wave is shown. Periodic wave solutions of the model system of equations are studied and it is shown that the waves that
are consequence of the irregular heart pulsations may be modeled by a sequence
of parts of such periodic wave solutions.
\end{abstract}
\section{Introduction}
Nonlinear phenomena are usual for physics and especially for fluid mechanics \cite{u1} - \cite{u4}. 
One of the most interesting nonlinear phenomena are the nonlinear
waves propagating in various media \cite{u5} - \cite{ux9} and especially in fluids \cite{u10} - \cite{u12}. 
In this paper we shall discuss nonlinear waves
connected to blood flow in large arteries \cite{u13} - \cite{u15}.
There exist many kinds of possible methodologies to investigate traveling
waves of the model nonlinear partial differential equations. One of them is to 
reduce the corresponding nonlinear PDE to a system of nonlinear ordinary 
differential equations  and then to investigate the obtained system
numerically. Another is to apply methods for obtaining 
(exact) solutions of the studied nonlinear partial differential equation \cite{k1} - \cite{v5}. We shall use the first of the above
methodologies in this paper. 
\par
Fluid flow connected to spreading of pressure pulses in elastic tubes is of interest for arterial mechanics of large arteries \cite{d1} - \cite{fu}. 
In this case the nonlinearities are important for the study of the flow and
because of this one has to use non-linear model differential equations.
In the mathematical models arteries usually are treated as circularly cylindrical long homogeneous isotropic tubes which length is much larger that 
its radius (i.e. the corresponding tube can be considered as thin tube with respect to the ratio length/thickness).
Below we shall study the propagation of non-linear waves in a fluid-filled long
elastic tube with variable radius.  The fluid will be incompressible (this a reasonable 
assumption for the case of blood flow in large arteries) and the 
tube will be isotropic, inhomogeneous and prestressed. The model equations of the flow will be reduced to a variant of the Korteweg-deVries equation with one 
variable coefficient. This nonlinear partial differential equation will be further reduced to a 
system of 3 nonlinear ordinary differential equations for the case of traveling waves. 
The system of nonlinear ordinary differential equations will be studied numerically.
\par
The organization of the paper is as follows. In Sect 2. we discuss the model
equations for the blood flow in a large artery. In Sect.3 by application of the
long-wave approximation the model equation will be reduced to a forced Korteweg-deVries equation and this equation will be further reduced to a system of 3 nonlinear ordinary differential equation. In Sect. 4 we perform a numerical
study of the system of nonlinear ODEs. Several concluding remarks are summarized in Sect. 5.  
\section{Mathematical formulation of the model}
For our study we shall use the nonlinear model presented in \cite{d1}. The model has two parts: equations of the elastic tube and
equation of the fluid in the tube. First we describe the model equations of
the elastic tube.
We set a cylindrical polar co-ordinate system with axial axis coinciding to the axis of the (straight) studied tube. We denote the
base vectors of this co-ordinate system as $\vec{e}_r$, $\vec{e}_\theta$, $\vec{e}_z$.
$R_0$ is the radius of the tube before the stretching. The tube is
assumed to be tapered (with small tapering angle $\Phi$). 
$Z^*$ is the axial co-ordinate along the axis of the elastic tube.
Because of the tapering the radial coordinate of the tube point with axial 
co-ordinate $Z$ will be $R_0 + Z^* \sin \Phi \approx R_0 + Z^* \Phi$. Thus the
initial coordinate of the point will be $\vec{R} = (R_0 + Z^* \Phi) \vec{e}_r
+ Z^* \vec{e}_z$.
In this case of initial conditions (with no stretching) the length of the
elements of the meridional and circumferential curves at the point with coordinate $\vec{R}$ are as follows.
The meridional curve  is a straight line. If the change of the $Z^*$
co-ordinate is $dZ^*$ then the change of the radius is $dR = dZ^* \sin \Phi
\approx dZ^* \Phi$. Then the length of the element of the meridional curve will
be $dS_Z^2 = {dZ^*}^2 + {dZ^*}^2 \Phi^2$ and then
\begin{equation}\label{ec1}
dS_Z = (1 + \Phi^2)^{1/2} dZ^*
\end{equation}
The length of the circumferential curve is given by the relationship
$dS_\theta/R = \sin (d \theta) \approx d \theta$ where 
$R = R_0 + Z^* \Phi$. Then
the length of the circumferential curve is
\begin{equation}\label{ec2}
dS_\theta = (R_0 + Z^* \Phi) d \theta
\end{equation}
\par
We assume that there is axial stretch of the tube. After the stretching the axial co-ordinate becomes
$z^* = \lambda_z Z^*$ where $\lambda_z$ is the axial stretch ratio (for the case without stretching $\lambda_z=1$).
In addition there is axially-dependent static pressure $P_0(Z^*)$ imposed on the tube 
(it has to be understood as the end of the diastolic pressure).
Let $r_0$ be the deformed radius of the tube at the origin of coordinate system 
Then the coordinate of
a point of the tube is
\begin{equation}\label{ax1}
\vec{r}_0 (z^*)= [r_0+ \phi z^* ]\vec{e}_r + z^* \vec{e}_z 
\end{equation}
where $\phi$ is the tapering angle after imposing the pressure (the tube
responds to this pressure by changing its form and Eq. (\ref{ax1}) is for the
case when the form of the tube after the imposing of the pressure remains cone).
If the form of the tube do not remains cone  then we have to introduce the 
function $f^*(z)$ which characterizes the radius change and instead of
Eq.(\ref{ax1}) we shall have
\begin{equation}\label{a1}
\vec{r}_0 (z^*)= [r_0+f(z^*)]\vec{e}_r + z^* \vec{e}_z 
\end{equation}
Now the lengths of the elementary meridional and circumferential curve elements 
(for the case when the deformed form of the tube doesn't remain cone) are as
follows. The length of the meridional curve element is:
\begin{equation}\label{ecxx1}
ds_z^0 = \left[1 + \left( \frac{\partial f^*}{\partial z^*} \right)^2 \right]^{1/2} dz^*
\end{equation}
The length of the curcumferential curve element is:
\begin{equation}\label{ecxx2}
ds_\theta^0 = (r_0 + f^*) d \theta
\end{equation}
\par
Additional (dynamical) deformation of the tube arises from the presence 
of fluid (blood) flow. This deformation depends on the coordinate
$z^*$ and on the time $t^*$. Let us denote this deformation as $u^*(z^*,t^*)$. 
Then the coordinate of the point of the tube becomes
\begin{equation}\label{a2}
\vec{r}_0 (z^*)= [r_0+f(z^*)+u^*(z^*,t^*)]\vec{e}_r + z^* \vec{e}_z 
\end{equation}
In this case the lengths of the elementary meridional and circumferential tubes are as follows. The length of the meridional curve element is:
\begin{equation}\label{ecyy1}
ds_z^0 = \left[1 + \left( \frac{\partial f^*}{\partial z^*}  + \frac{\partial u^*}{\partial z^*}\right)^2 \right]^{1/2} dz^*
\end{equation}
The length of the circumferential curve element is:
\begin{equation}\label{ecyy2}
ds_\theta^0 = (r_0 + f^* + u^*) d \theta
\end{equation}
\par
The corresponding stretch ratios are
\begin{equation}\label{st1}
\lambda_1 = \frac{d s_z}{d S_Z}; \ \ \lambda_2 = \frac{d s_\theta}{d S_\theta}
\end{equation}
After the static and dynamic deformation (when the tube does not  remains cone after the static deformation)
\begin{equation}\label{st5}
\lambda_1 = \lambda_z \frac{[(1 + [(\partial f^* / \partial z^*) + (\partial u^*/\partial z^*)]^2]^{1/2}}{(1+\Phi^2)^{1/2}}; \ \  \ 
\lambda_2 = \lambda_z \frac{r_0+ f^* + u^* }{\lambda_z R_0 + z^* \Phi}
\end{equation}
For the case when $\Phi=0$  (i.e.the tube before applying the static pressure is cylinder and not a cone):
\begin{equation}\label{st6}
\lambda_1 = \lambda_z [(1 + [(\partial f^* / \partial z^*) + (\partial u^*/\partial z^*)^2]^{1/2}]; \ \  \ 
\lambda_2 = \lambda_z \frac{r_0+ f^* + u^* }{\lambda_z R_0}
\end{equation}
The relationship for $\lambda_2$ can be written as follows
\begin{equation}\label{st7}
\lambda_2 = \frac{r_0}{R_0} + \frac{f^* + u^*}{R_0} = \lambda_\theta + \frac{f^* + u^*}{R_0}
\end{equation}
where $\lambda_\theta = r_0/R_0$. 
\par
In this case the tube is deformed and in the general case the unit normal vector $\vec{n}$ does not
coincide to the unit vector $\vec{e}_r$. The unit tangential vector $\vec{t}$ of the curved surface also
do not coincide to the vector $\vec{e}_z$. The unit normal and tangential vectors can be expressed by
the unit vectors $\vec{e}_r$, $\vec{e}_z$. When we take into account that
\begin{equation}\label{st8}
[(1 + [(\partial f^* / \partial z^*) + (\partial u^*/\partial z^*)]^2]^{1/2} = \frac{\lambda_1}{\lambda_z}
\end{equation}
then
\begin{equation}\label{st9}
\vec{t} =\frac{[(\partial f^* / \partial z^*) + (\partial u^*/\partial z^*)] \vec{e}_r + 
\vec{e}_z}{(1 + [(\partial f^* / \partial z^*) + (\partial u^*/\partial z^*)]^{2})^{1/2}}  = \frac{\lambda_z}{\lambda_1} ([(\partial f^* / \partial z^*) + (\partial u^*/\partial z^*)]\vec{e}_r + \vec{e}_z)
\end{equation}
\begin{equation}\label{st10}
\vec{n} = \frac{\vec{e}_r - [(\partial f^* / \partial z^*) + (\partial u^*/\partial z^*)] \vec{e}_z}{(1 + [(\partial f^* / \partial z^*) + (\partial u^*/\partial z^*)]^{2})^{1/2}}  = \frac{\lambda_z}{\lambda_1}
[\vec{e}_r - [(\partial f^* / \partial z^*) + (\partial u^*/\partial z^*)] \vec{e}_z]
\end{equation}
\par
4 forces are responsible for the movement of a fluid element.
The first force is the force of movement in radial direction due to the 
existing pressure difference. The second force is the force along the meridional curve. The third force is the force acing along the circumferential curve.
The force connected to the movement of the tube element is equal to the mass of the tube element multiplied by its acceleration. Let the tube thickness before 
the static deformation be $H$. The thickness of the tube after the static deformation will be $h$. Then the mass of the tube element is approximately 
$\rho_w h R d\theta dz$ where $\rho_w$ is the mass density of the material of 
the tube. The acceleration of the element is equal  to $\frac{\partial^2 
u^*}{\partial {t^*}^2}$. Thus this term of the equation of balance of forces 
becomes  $\rho_w h R d\theta dz \frac{\partial^2 u^*}{\partial {t*}^2}$. The tube thickness $h$ after the static deformation can be expressed by the tube 
thickness $H$ before the static deformation. The assumption is that the 
material is incompressible. This leads to
\begin{equation}\label{ex1}
\lambda_1 \lambda_2 h= H \to h= \frac{H}{\lambda_1 \lambda_2}
\end{equation}
\par
In our case (when axial stretching exists) the initial radius $R_0$
for tube with length $dl$ transforms to radius $R$ for a tube of length $\lambda_z dl $. Assuming that the area remains the same we obtain $R=R_0/\lambda_z$ Thus for the force per unit $d \theta dz$ we obtain
$\rho_w H (R_0/\lambda_z) \frac{\partial^2 u^*}{\partial {t*}^2}$.
\par
The second force is the pressure force that acts on the tube element. This 
force is equal to the  pressure difference $P-P_e$ (where $P$ is the pressure in the tube and $P_e$ is the external pressure) multiplied by the  surface of the tube element that is $R d \theta dz $ . Then this force becomes 
 $(P-P_e) \cos \phi (R d \theta d l)$. In our case $P_e = P_0$ and $P-P_e = P^*$. The co-ordinate $R$ connected to the element of the tube is $R=r_0+f^* +
 u^*$. And the entire force per unit $d \theta d z$ becomes  $P^*(r_0+f^* + u^*)$.
\par
The remaining two forces are  connected to the membrane forces $T_1$ and  $T_2$ that act along the circumferential and meridional curves of the tube. For
unit $d \theta dz$ the force acting along the meridional curve is $F_2 \vec{t}$. Its vertical component is $T_2 = F_2 \vec{t} \cdot \vec{n}$ from where $T_2 =  - F_2 \lambda_z /\lambda_1$ and $F_2 = - T_2 \lambda_1/\lambda_z$.
The force $F_1$ is \cite{d1, goldenvizer}
 \begin{equation}\label{ex2}
 F_1 = \frac{\partial}{\partial z^*} \bigg[\frac{\lambda_z}{\lambda_1} (r_0+f^*+u^*) \bigg(\frac{\partial f^*}{\partial z} + \frac{\partial u^*}{\partial z} \bigg) T_1 \bigg]
\end{equation}
\par
The balance of the above 4 forces is
 \begin{eqnarray}\label{a3}
 -\frac{\lambda_1}{\lambda_z} T_2 +  \frac{\partial}{\partial z^*}
 \Bigg \{ \frac{\lambda_z (r_0 + f^* + u^*) (\frac{\partial f^*}{\partial z^*} + \frac{\partial u^*}{\partial z^*})}{\lambda_1} T_1 \Bigg \}
 + \nonumber \\
 P^*(r_0+f^*+u^*) = \rho_w H \frac{R_0}{\lambda_z} \frac{\partial^2 u^*}{\partial {t^*}^2}
 \end{eqnarray}
\par
Let $\mu(z^*)$ be the variable shear modulus of the tube material 
and  $\mu(z^*) \Pi$ be the strain energy function of the membrane. Then the membrane forces $T_{1,2}$ can be written as
\begin{equation}\label{a4}
T_1 = \frac{H}{\lambda_2} \mu(z^*) \frac{\partial \Pi}{\partial \lambda_1}; \ \
T_2 = \frac{H}{\lambda_1} \mu(z^*) \frac{\partial \Pi}{\partial \lambda_2}
\end{equation}
After a substitution of Eq.(\ref{a4}) in Eq. (\ref{a3}) we obtain the pressure
$P^*$ as a function of $u^*$ and its derivatives.
\begin{eqnarray}\label{a5}
P^* &=& \frac{\rho_w H R_0}{\lambda_z (r_0 + f^* + u^*)} \frac{\partial^2 u^*}{\partial {t^*}^2} + \frac{\mu(z^*)}{\lambda_z(r_0+f^*+u^*)} \frac{\partial \Pi}{\partial \lambda_2} - \nonumber \\
&& \frac{\lambda_z R_0}{(r+_0+f^*+u^*)} \frac{\partial}{\partial z^*} \bigg[ \frac{\mu(z^*)}{\lambda_1} \bigg( \frac{\partial f^*}{\partial z^*} + \frac{\partial y^*}{\partial z^*}\bigg)
\frac{\partial \Pi}{\partial \lambda_1} \bigg]
\end{eqnarray}
\par
The model equations of the fluid in the tube are as follows.
The blood in a large arteries can be approximated by a Newtonian fluid with
respect to its flow (this is not the case for blood flow in small arteries).
In addition the viscosity of the blood may be neglected as s first approximation
\cite{fung, rudinger} and the variation of the quantities with the radial coordinate will be disregarded too. What remains from the averaged Navier-Stokes equations in cylindrical coordinates is  
\begin{equation}\label{a6}
\frac{\partial w^*}{\partial t^*} + w^* \frac{\partial w^*}{\partial z^*} +
\frac{1}{\rho_f} \frac{\partial P^*}{\partial z^*} =0
\end{equation}
where $\rho_f$ is the density of the fluid and $W^*$ is the axial fluid
velocity. In addition
\begin{equation}\label{a7}
\frac{\partial A^*}{\partial t^*} + \frac{\partial}{\partial z^*} (w^* A^*) =0
\end{equation}
where $A^*$ is the cross-sectional area of the tube. This area is
\begin{equation}\label{a8}
A^* =  \pi (r_0 + f^* + u^*)^2
\end{equation}
\section{Non-dimensionalization of the equations and long-wave approximation}
The following nondimensional quantities $t$, $z$, $R_0$, $u$, $w$, $m$, $p$, $c_0$ and $E(z)$ are introduced as follows
\begin{eqnarray}\label{a9}
t^* = t \left( \frac{R_0}{c_0} \right); \ c_0^2 = \frac{\mu_0 H}{\rho_f R_0};
\ z^* = R_o z; \ r_0 = \lambda_{\theta} R_0 \nonumber \\
u^* = R_0 u, m = \frac{\rho_w H}{\rho_f R_0}; \ w^* = w c_0;
P^* = p \rho_f c_0;  \mu = \mu_0 E(z)
\end{eqnarray}
The model equations for the unknown functions $u$, $w$ and $p$ in dimensionless coordinates are:
\begin{eqnarray}\label{a10}
\frac{\partial w}{\partial t} + w \frac{\partial w}{\partial z} + \frac{\partial p}{\partial z} = 0
\end{eqnarray}
\begin{eqnarray}\label{a11}
2 \frac{\partial u}{\partial z} + 2 w \left( \frac{\partial f}{\partial z} +
\frac{\partial u}{\partial z} \right) + (\lambda_{\theta} + f + u) \frac{\partial w}{\partial z} =0
\end{eqnarray}
\begin{eqnarray}\label{a12}
p &=& \frac{m}{\lambda_z (\lambda_{\theta} + f + u)} \frac{\partial^2 y}{\partial t^2} + \frac{E(z)}{\lambda_z (\lambda_{\theta} + f + u)} \frac{\partial \Pi}{\partial \lambda_2} - \nonumber \\
&& \frac{\lambda_z}{(\lambda_{\theta} + f + u)} \bigg[ \frac{E(z) \big( \frac{\partial f}{\partial z} +
\frac{\partial u}{\partial z}\big)}{\lambda_1} \frac{\partial \Pi}{\partial \lambda_1} \bigg]
\end{eqnarray}
In order to proceed further we shall consider the case of propagation of small
(but finite) amplitude waves in an inhomogeneous thin elastic tube of variable radius and filled with Newtonian fluid. We assume that $\epsilon$ is a
small parameter and introduce the following coordinates
\begin{equation}\label{a13}
\xi = \epsilon^{1/2} (z - gt); \ \ \tau = \epsilon^{3/2} z
\end{equation}
From here $z = \tau \epsilon^{-2/3}$ and we can use the notations $h(\epsilon,\tau) = f(z)$ and $\hat{E}(\tau, \epsilon) = E(z)$.
\par
The next step is to expand $u$, $w$, $p$, $h$ and $\hat{E}$ in series of
the small parameter $\epsilon$
\begin{eqnarray}\label{a14}
u &=& \epsilon u_1(\xi,\tau) + \epsilon^2 u_2(\xi,\tau)+ \dots \nonumber \\
w &=& \epsilon w_1(\xi,\tau) + \epsilon^2 w_2(\xi,\tau)+ \dots \nonumber \\
p &=& p_0 +  \epsilon p_1(\xi,\tau) + \epsilon^2 p_2(\xi,\tau)+ \dots \nonumber \\
h(\epsilon,\tau) &=& 1 + \epsilon h_1(\tau) + \dots \nonumber \\
\hat{E}(\epsilon,\tau) &=& 1 + \epsilon E_1(\tau) + \dots \nonumber \\
\end{eqnarray}
Let $U(\xi, \tau) = u_1$. From the systems of equations of orders $\epsilon$ and $\epsilon^2$ we obtain for $U$ the partial differential equation
\begin{equation}\label{a15}
\frac{\partial U}{\partial \tau} + \mu_2 U \frac{\partial U}{\partial \xi} +
\mu_2(\tau) \frac{\partial U}{\partial \xi} + \mu_3 \frac{\partial^3 U}{\partial \xi^3} = 0
\end{equation}
The other unknown functions are
\begin{eqnarray}\label{a16}
w_1 &=& 2 \frac{g}{\lambda_0}[U + \hat{w}_1(\tau)] \nonumber \\
\hat{w}_1(\tau) &=& - \left( h_1 + \frac{\beta_0}{\beta_1} E_1 \right) \nonumber \\
\beta_0 &=& \frac{1}{\lambda_{\theta} \lambda_z} \frac{\partial \Pi}{\partial \lambda_{\theta}} \mid_{u=0}; \ \beta_1 = \frac{1}{\lambda_{\theta} \lambda_z} \frac{\partial^2 \Pi}{\partial \lambda_{\theta}^2} \mid_{u=0} - \frac{\beta_0}{\lambda_{\theta}} \nonumber \\
g^2 &=& \frac{\beta_1}{2 \lambda_{\theta}} \nonumber \\
p_1 &=& 2 \frac{g^2}{\lambda_{\theta}}(h_1 + U) + \beta_0 E_1
\end{eqnarray}
and $\mu_{1,2,3}$ are as follows
\begin{eqnarray}\label{a17}
\mu_1 &=& \frac{5}{2 \lambda_{\theta} + \frac{\beta_2}{\beta_1}} \nonumber \\
\beta_2 &=& \frac{1}{2 \lambda_{\theta} \lambda_z} \frac{\partial^3 \Pi}{\partial \lambda_{\theta}^3} \mid_{u=0} + \frac{\beta_0}{\lambda_{\theta}^2} - \frac{\beta_1}{\lambda_{\theta}} \nonumber \\
\mu_2(\tau)&=& \left( \frac{\beta_2}{\beta_1} -  \frac{3}{2 \lambda_{\theta}}\ \right) h_1(\theta) + \left( \frac{1}{2} - \frac{2 \beta_0}{\beta_1 \lambda_{\theta}} E_1(\tau) \right)  \nonumber \\
\mu_3 &=& \frac{1}{\lambda_z \lambda_{\theta}} \left(\frac{m}{4 \lambda_{\theta}} - \frac{\alpha_0}{2 \beta_1} \right)
\end{eqnarray}
\par
Finally we have to deal with the variable coefficient $\mu_2(\theta)$ in Eq.(\ref{a15}). We introduce the new coordinate
\begin{equation}\label{a18}
\eta = \xi + \tau - \int_{0}^{\tau}ds \ \mu_2(s) 
\end{equation}
The substitution of Eq.(\ref{a18}) in Eq.(\ref{a15}) lead to the equation
\begin{equation}\label{a19}
\frac{dU}{d \eta} + \mu_1 U \frac{dU}{d \eta} + \mu_3 \frac{d^3U}{d \eta^3} =0
\end{equation}
Let $V = dU/d \eta$ and $W = dV/d \eta$. Then Eq.(\ref{a19}) is reduced 
to the following system of 3 equations for the unknown functions $U,V,E$:
\begin{eqnarray}\label{a20}
\frac{dU}{d \eta} &=& V, \nonumber \\
\frac{dV}{d \eta} &=& W, \nonumber \\
\mu_3 \frac{dW}{d \eta} &=& -V(1+ \mu_1 U)
\end{eqnarray}
We remember that $U$ is connected to the deformation of the tube due to the
presence of fluid and this quantity is the main quantity of interest for us
in this paper.
\section{Numerical results}
Eq.(\ref{a15}) possesses a solitary wave solution. This solitary wave solution
is connected to the solitary wave solution of the classical Korteweg-deVries
equation
\begin{equation}\label{b1}
\frac{\partial A}{\partial t} + \alpha A \frac{\partial A}{\partial x} +
\beta \frac{\partial^3 A}{\partial x^3} =0
\end{equation}
Now let
\begin{equation}\label{b2}
A^*=\alpha A /6; \ \ \ x^* = x/\beta^{1/2}; \ \ \  t^* = t / \beta^{1/2}
\end{equation}
The result of substitution of Eq.(\ref{b2}) in Eq.(\ref{b1}) (we drop the $^*$-s)
is
\begin{equation}\label{b3}
\frac{\partial A}{\partial t} + 6 A \frac{\partial A}{\partial x} +
\frac{\partial^3 A}{\partial x^3} =0
\end{equation}
Let us search for travelling-wave solutions of Eq.(\ref{b3}) of the kind 
$A(x,t)=A(\zeta)=A(x - vt)$. Eq.(\ref{b3}) becomes
\begin{equation}\label{b4}
\frac{d A}{d \zeta} - \frac{6}{v} A \frac{d A}{d \zeta}- 
\frac{1}{v} 
\frac{d^3 A}{d \zeta^3} =0
\end{equation}
which is the same as Eq.(\ref{a19}) when $\mu_1 = - 6/v$ and $\mu_3 = -1/v$.
The solitary wave solution of Eq.(\ref{a16}) is
\begin{equation}\label{b5}
U(\eta) = \frac{v}{2} \textrm{sech}^2 \left( \frac{v^{1/2}}{2} \eta \right) 
\end{equation}
The realization of this solution for the case of blood flow in large
arteries however has low probability because of two reasons. First the
existence of the solution (\ref{b5}) requires a relationship between $\mu_1$
and $\mu_3$ (namely $\mu_1 = 6 \mu_3$) that may not be present in the
practical situations. And second the realization of the solution (\ref{b5})
requires specific boundary conditions. The derivatives of (\ref{b5}) are
as follows
\begin{eqnarray}\label{b6}
\frac{dU}{d \eta} = - \frac{v^{3/2} \sinh (v^{1/2} \eta/2)}{\cosh^3(v^{1/2} \eta/2)}; \ \frac{d^2U}{d \eta^2} = \frac{v^2[2 \cosh^2(v^{1/2} \eta/2)-3]}{4
\cosh^4(v^{1/2} \eta/2)} \nonumber \\
\frac{d^3 U}{d \eta^3} = - \frac{v^{5/2} \sinh(v^{1/2} \eta/2)[\cosh^2(v^{1/2} \eta/2)-3]}{2 \cosh^5(v^{1/2} \eta/2)}; \ \dots
\end{eqnarray}
Thus the boundary conditions for realization of the solitary wave at $\eta=0$
should be 
\begin{equation}\label{b7}
U(0) = \frac{v}{2}; \frac{dU}{d \eta}\mid_{\eta=0} = 0; \frac{d^2U}{d \eta^2}\mid_{\eta=0} = - \frac{v^2}{4}; \frac{d^3 U}{d \eta^3}\mid_{\eta=0} = 0;
\dots 
\end{equation}
The realization of these boundary conditions is not very probable as the
heart pulsations are slightly irregular with respect to amplitude and time
between the beats. Thus if a solitary wave solution is realized for a pulsation
the next pulsation will lead to slight change of the boundary conditions and
the next wave will be not solitary. Then another scenario for blood waves is
more probable and this scenario is connected to the periodic solutions of
Eq.(\ref{a19}).
\par
The periodic wave solutions of Eq.(\ref{a19}) can be realized for much more
values of the boundary conditions and for different amplitudes of the blood
waves. Several examples of periodic solutions of Eq.(\ref{a19}) obtained through
the system of equation (\ref{a20}) are shown in Fig. 1.
\vspace{\baselineskip}
\par
\begin{figure}[ht!]
\centering
\includegraphics[scale=0.45]{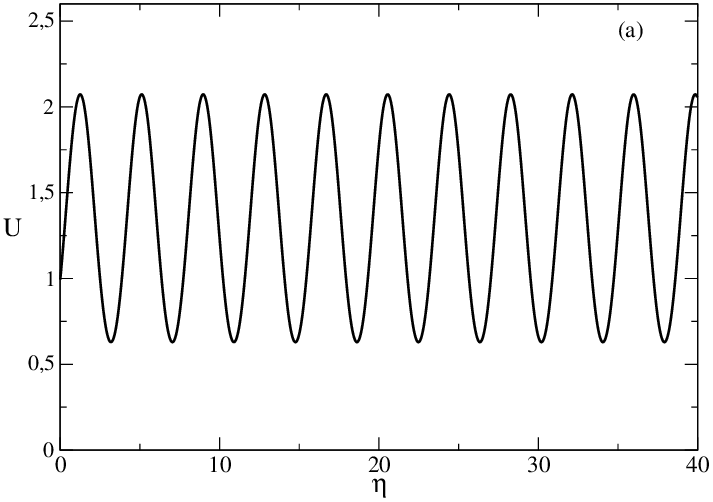} \hspace{0.2cm}\vspace{0.6cm}
\includegraphics[scale=0.45]{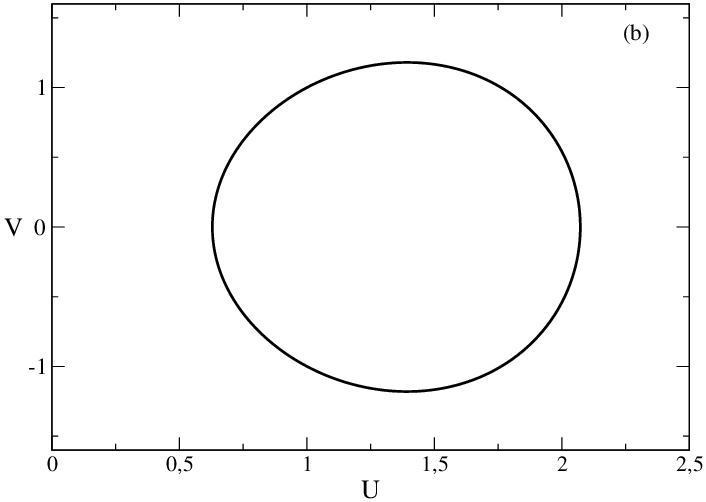}
\vspace{\baselineskip}
\includegraphics[scale=0.45]{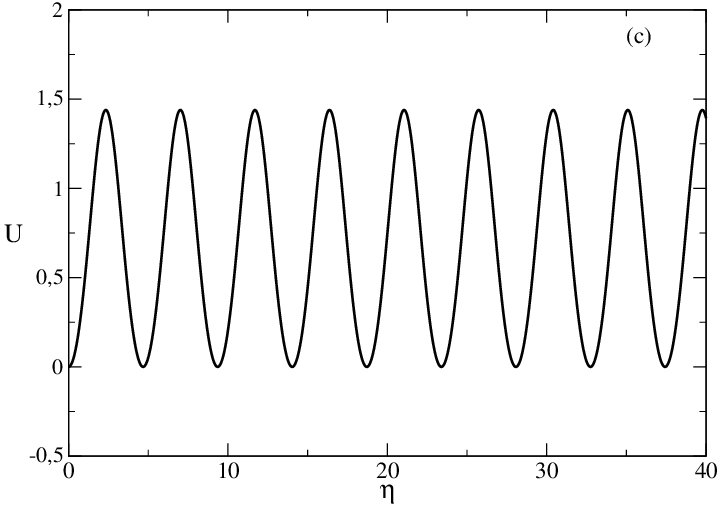}
\caption{Examples for periodic wave solutions of Eq.(\ref{a20}). Figure (a):
$\mu_1 = 1.43$; $\mu_3 = 1.1$. $U(0)=V(0)=W(0)=0.99$. Figure (b): Diagram of the components $U$ and $V$ for the solution with parameters and initial conditions
that are the same as those in Figure (a). Figure(c): $\mu_1 = 4.0$; $\mu_3 = 2.1$. $U(0)=V(0)=0$; $W(0)=1$. One may observe that the period and the amplitude of this solution are different with respect to the period and amplitude of the
solution from Figure (a).}
\end{figure}
Because of their larger parameter regions of existence the periodic solutions can be used for construction of the wave motion in the blood in presence of
irregularity of heart beats as follows. Let the heart makes a pulsation and the
blood wave start to propagate in the artery. This can be modelled by a half a
period of the periodic wave solution of Eq.(\ref{a19}). When the next pulsation comes one can stop at the corresponding values of $U$ and its derivatives and can treat them as the new initial conditions. These new initial conditions describe slightly different periodic wave. Half a period of this wave can be used to model the blood flow wave up to the moment of the next pulsation. At this moment the reached values of $U$ and its derivatives are again the initial
conditions that describe the blood wave corresponding to the third pulsation, etc. In such a way a sequence of slightly different waves (shown in Fig.2.)
\vspace{0.55cm}
\begin{figure}[ht!]
\centering
\includegraphics[scale=0.7]{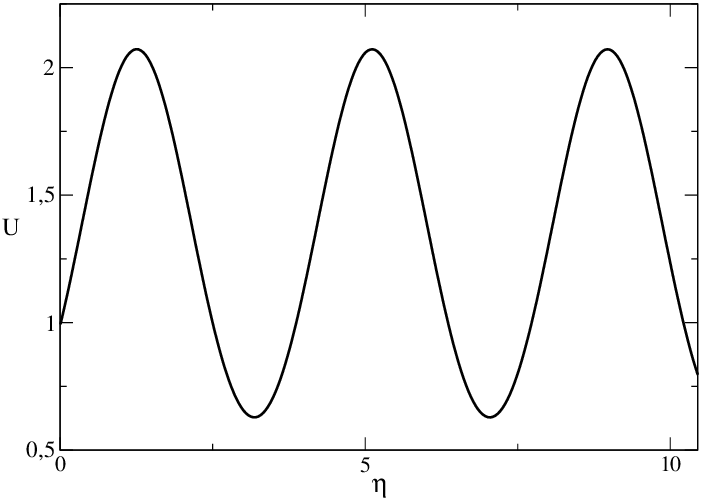}
\caption{Periodic blood waves constructed by parts of periodic solutions of
Eq.(\ref{a19}) by the algorithm described in the text. It may seem that a single wave is shown at the figure. Actually the
wave is constructed by the parts of 3 different periodic solutions of Eq.(\ref{a19}). Namely  it is assumed that the first pulsation started at $\eta=0$ with initial
conditions $U(0)=V(0)=W(0)=0.99$ and $\mu_1 = 1.43$; $\mu_3 = 1.1$. The second pulsation was at $\eta=2.844$ with
initial conditions $U(2.844)=0.7253$; $V(2.844)=-0.5571$; $W(2.844)=1.5255$ and
$\mu_1 = 1.42$; $\mu_3 = 1.11$. The
third pulsation was at $\eta=6.924$ with initial conditions $U(6.924)=0.6409$; $V(6.924)=-0.2000$; $W(6.924)=1.6772$ and $\mu_1 = 1.44$; $\mu_3 = 1.08$. }
\end{figure}
may model the slight irregularity of the heart activity. It is well known that the time intervals between the heart pulsations are long-range correlated
and this is one of the numerous arising of long-range correlations in various systems \cite{ph1} - \cite{corv2}. Such long-range correlated pseudorandom sequences modeling
heart activity may be generated by a computer program and the values in the sequence will determine the end of the corresponding wave train and the beginning of the next wave train. This was realized in Fig.2.
\section{Concluding remarks}
In this paper we have shown that the area of research on blood flow is a
large area for application of the methods of nonlinear dynamics. Even the
relative single problems such as investigation of waves in large arteries
(where the fluid can be treated as Newtonian and the long-wave approximation
significantly simplifies the equations) lead to relatively complicated model
equations such as the discussed above variable coefficient KdV equation.
We have stressed that the solitary wave solution of the model equation requires very specific initial conditions and relationship among the two model parameters. Because of this the probability for realization of this solution is small. We have discussed another kind of solutions of the nonlinear model
equation that are much more probable for realization: they do not require relationship between the two parameters of the model and are robust against change of the initial conditions due to the irregularities of the pulsation dynamics of the heart. These solutions are the periodic solutions of the system of equations (\ref{a20}). Using parts of these solutions one can construct a model profile of a blood waves that reflect the irregularities and the long-range correlations presented in the pulsation activity of human heart.
\section{Asknowledgment}
This research was supported partially by the project FNI I 02/53 "Computer modeling and
clinical study of arterial aneurysms of humans".

\end{document}